\documentclass[conference]{IEEEtran}
\IEEEoverridecommandlockouts
\usepackage{amsmath,amssymb,amsfonts}
\usepackage{algorithmic}
\usepackage{graphicx}
\usepackage{textcomp}
\usepackage{xcolor}
\usepackage{hyperref} 
\usepackage{setspace}

\usepackage[
backend=biber,
style=ieee,
]{biblatex}
\makeatletter \def\@IEEEpubidpullup{8\baselineskip} \makeatother 

\usepackage{fancyhdr}
\usepackage{kantlipsum}
\fancypagestyle{plain}{
\fancyhf{}
\fancyhead[C]{Conference on \LaTeX} 
\fancyfoot[L]{This is a notice}
} \usepackage{eso-pic}

\begin{document}

\title{Graph Learning for Bidirectional Disease Contact Tracing on Real Human Mobility Data}

\author{\IEEEauthorblockN{ Sofia Hurtado}
\IEEEauthorblockA{\textit{Dept. of Electrical Computer Engineering} \\
\textit{The University of Texas at Austin}\\
Austin, TX, 78712, USA \\
slhurtad@utexas.edu}
\and
\IEEEauthorblockN{Radu Marculescu}
\IEEEauthorblockA{\textit{Dept. of Electrical Computer Engineering} \\
\textit{The University of Texas at Austin}\\
Austin, TX, 78712, USA \\
radum@utexas.edu}
}
\maketitle

\begin{figure*}[h]
\centering
\includegraphics[width= .8 \textwidth]{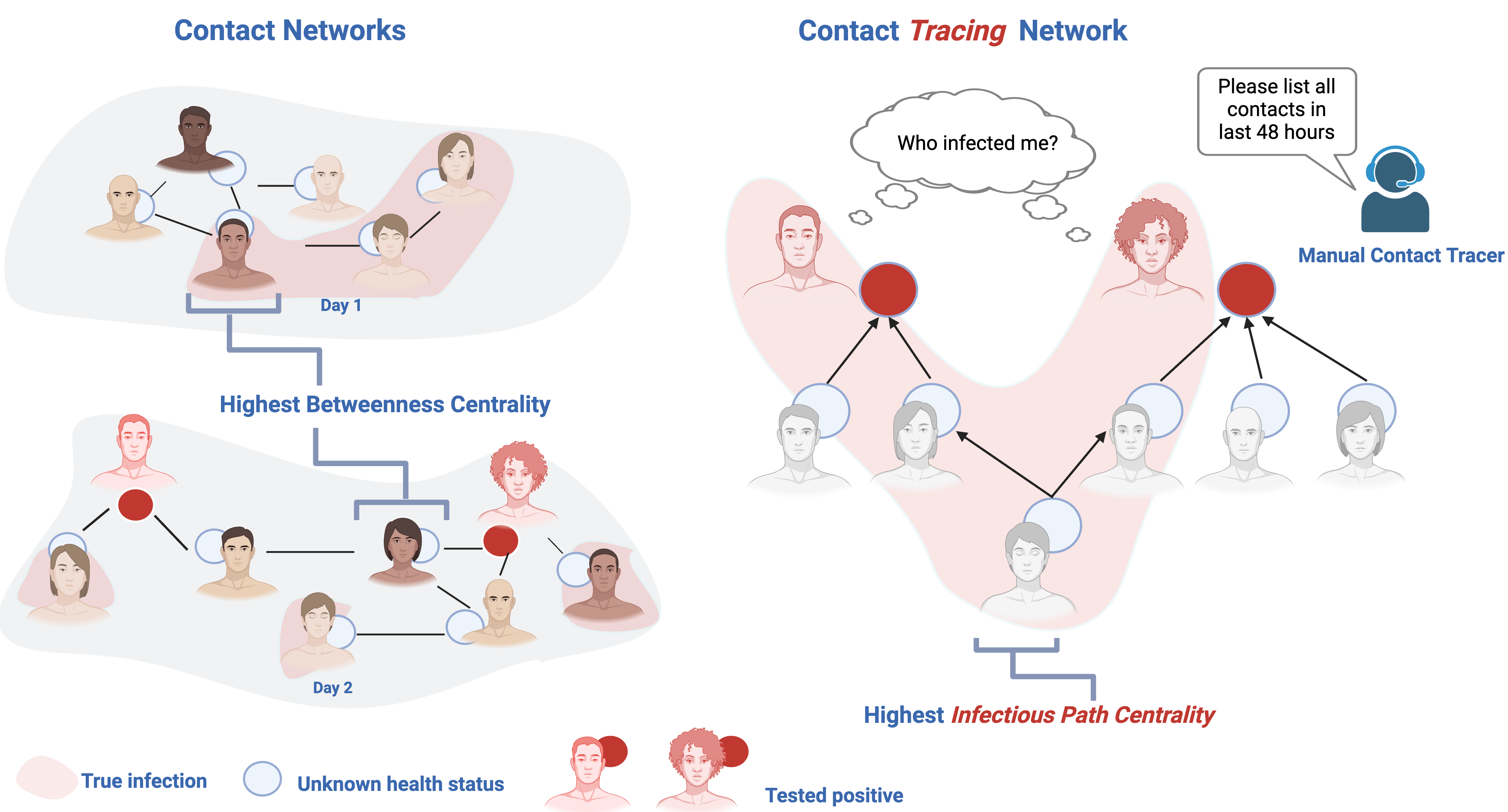}
\caption{   
Manual contact tracing involves collecting past interactions to construct a directed acyclic graph (DAG), where parent nodes are potential sources of infection for their child nodes (forming a contact tracing network). When identifying superspreaders, potential infections, or vaccination candidates, most studies use network analysis techniques such as betweenness centrality on networks with static interactions but dynamic node labels (i.e., health status). However, as illustrated in the contact networks from Day 1 and Day 2 (left), nodes with the highest betweenness centrality \cite{newman_centrality} do not necessarily hold significant roles in the contact tracing network (right). Instead, nodes with the highest value of our proposed metric, Infectious Path Centrality—which measures the number of paths connecting two positive leaf nodes—are often the most recent common ancestors, making them (and their offspring) crucial for targeted quarantines. We evaluate our metric by comparing its effectiveness in a bidirectional graph learning mitigation framework, which uses this new transmission network metric to identify and quarantine unseen branches of the disease, against traditional forward contact tracing that quarantines those who test positive.
 }
\label{fig1}
\end{figure*}

\begin{abstract}
For rapidly spreading diseases where many cases show no symptoms, swift and effective contact tracing is essential. While exposure notification applications provide alerts on potential exposures, a fully automated system is needed to track the infectious transmission routes. To this end, our research leverages large-scale contact networks from real human mobility data to identify the path of transmission. More precisely, we introduce a new Infectious Path Centrality network metric that informs a graph learning edge classifier to identify important transmission events, achieving an F1-score of 94\%. Additionally, we explore bidirectional contact tracing, which quarantines individuals both retroactively and proactively, and compare its effectiveness against traditional forward tracing, which only isolates individuals after testing positive. Our results indicate that when only 30\% of symptomatic individuals are tested, bidirectional tracing can reduce infectious effective reproduction rate by 71\%, thus significantly controlling the outbreak.
\end{abstract}


\section{Introduction}

Contact tracing has long been a cornerstone public health strategy for managing outbreaks of highly traceable diseases such as Ebola and Rabies \cite{r1}\cite{r2}. In such instances, the transmission routes are typically clear, as infectees can recall specific events like animal bites or direct contact with infectious individuals. However, managing pathogens with aerosol transmission and asymptomatic cases like SARS-CoV-2 \cite{r3_full}, RSV \cite{r4}, or Influenza \cite{r5} presents greater disease containment challenges, as seen during the COVID-19 pandemic.

To contain a disease outbreak, manual contact tracers need to contact each person who tests positive to identify their recent contacts from the past 24-48 hours \cite{r6}. However, the effectiveness of manual tracing depends on the accuracy of people's memories, their awareness of their surroundings, and the tracer's ability to reach out to these contacts \cite{r7}. 

In the beginning of an outbreak, the goal of manual contact tracers is to determine the initial source, often referred to as "patient zero." However, as an outbreak escalates into an epidemic, manual tracing can become overwhelmed by rapidly spreading pathogens, especially when community spread occurs \cite{r8} \cite{r9}. This is because numerous undetected transmission events within the community can make it difficult for individuals to trace their infections to the original source. The goal of contact tracers then pivots to identifying untested paths of transmission to notify people who may be infectious. 

In response to the rapidly evolving SARS-CoV-2 virus, companies raced to digitize disease contact tracing. By monitoring interactions among individuals, modern systems can function as continuous disease surveillance tools. Exposure notification applications, for instance, alert users to recent contacts with infected individuals, leveraging advancements in location-tracking technology \cite{r10}. Despite these advances, a critical question remains: \textit{Can we effectively backtrack the chain of transmission, particularly for rapidly spreading diseases where many carriers are asymptomatic?}

With access to the Foursquare Mobility dataset \cite{r3_new} that contains visits at various Points of Interest (POIs), we present an \textit{always-ready} disease surveillance system that tracks the past interactions and performs retroactive disease path detection. In addition, we propose a new propagation network metric, \textit{Infectious Path Centrality}, that characterizes the centrality of a node along a \textit{path of a transmission} on a contact tracing network; this is in contrast to the static betweenness centrality on the instantaneous contact networks \cite{r28} (Fig. \ref{fig1}).

In this research, we have developed a graph learning framework that utilizes our newly proposed metric (Infectious Path Centrality) as node features. This framework automates the bidirectional contact tracing process, allowing it to identify and quarantine more individuals along the suspected transmission paths. To this end, our contributions are as follows:

\begin{enumerate}
   \item We introduce a novel Infectious Path Centrality metric, which measures how central a node is among all infectious individuals in a contact tracing network.
    \item We provide an automated bidirectional contact tracing graph learning framework using real mobility data that identifies transmission events along various paths.
    \item We show that bidirectional contact tracing is more effective than forward contact tracing at reducing the disease's effective reproduction rate. 
\end{enumerate}

Taken together, our contributions can help build the technology needed to mitigate the next unknown disease outbreak. The remainder of this paper is organized as follows: Section II discusses prior work, Section III describes our approach, Section IV presents our experimental results. Finally, Section V summarizes our contributions.

\begin{figure*}[ht]
\centering
\includegraphics[width=.9 \textwidth]{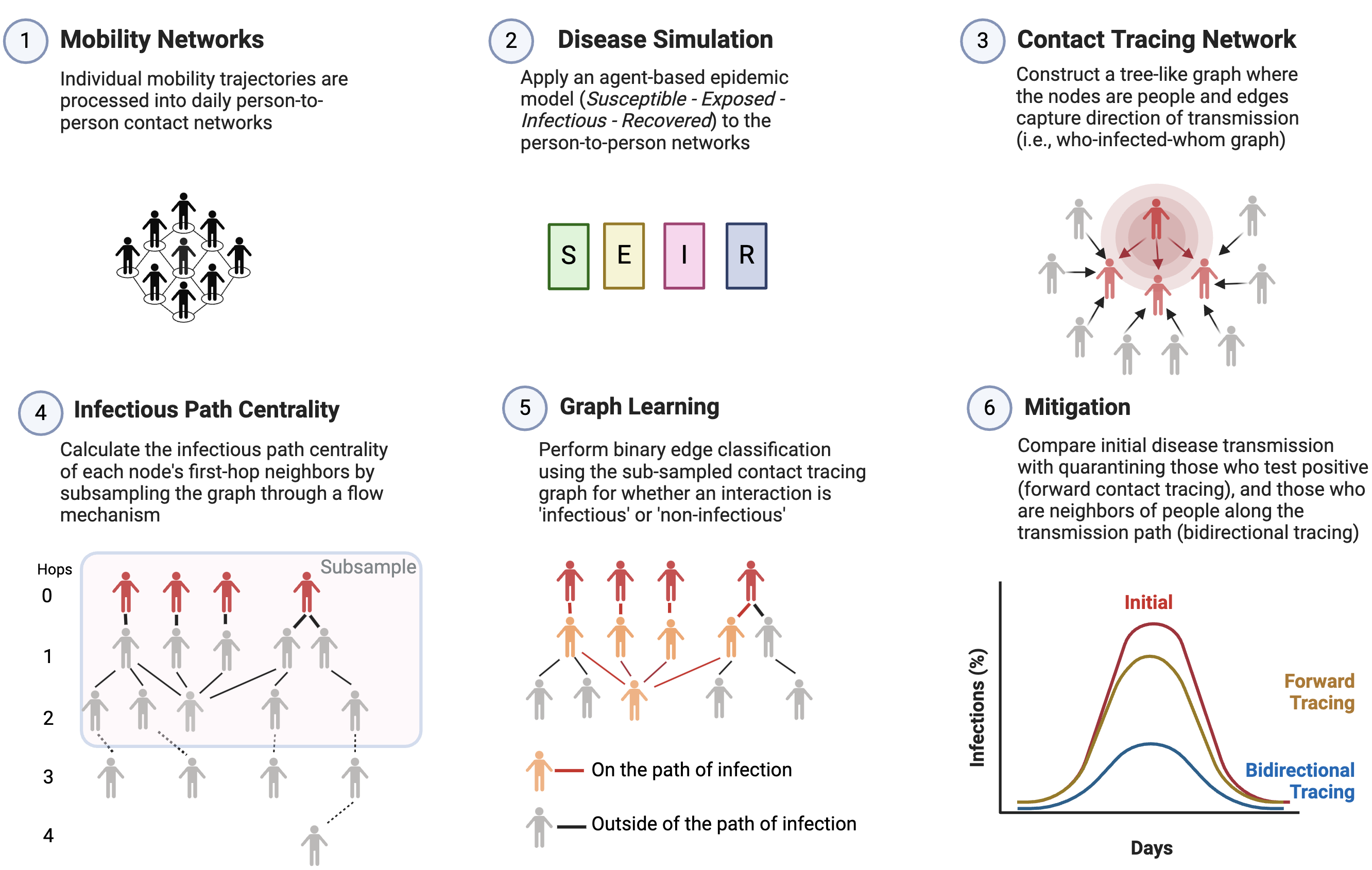}
\caption{   
Our experiment starts by processing Foursquare mobility data containing device dwell times at POIs into person-to-person contact networks (step 1). Two devices (i.e., people) are connected when they visit the same location within the same hour. We then apply an agent-based epidemiological SEIR (\textit{Susceptible}-\textit{Exposed}-\textit{Infectious}-\textit{Recovered}) model \cite{rSEIR} on the dynamic contact networks (step 2). To keep track of transmission events, we form a contact \textit{tracing} network where the parent is a potential source of infection to the child (step 3). For every transmission event (infection in step (2)), we add the infectious interaction to the contact tracing network, as well as all other interactions the infectee has on the day of infection. This mimics the manual contact tracing where someone recalls all of their interactions on the day they got infected. Step 4 consists of calculating our proposed Infectious Path Centrality metric that is then used as features in edge classification (step 5). The graph learning module classifies the edges in the contact tracing network as being 'infectious' (i.e., a true transmission event), or 'non-infectious). Finally, we test the efficacy of our approach by comparing a population seeded with the same infectious individuals that undergoes no mitigation, forward contact tracing, and bidirectional contact tracing using our mitigation framework (step 6). 
 }
\label{fig2}
\end{figure*}

\section{Prior Work and Novel Contribution}\label{sec2}
In this section we present the relevant prior work in disease contact tracing, targeting strategies, and graph learning. 

\subsection{Disease Contact Tracing}

Traditionally, contact tracing has been a manual process, relying heavily on interviews conducted by trained health workers to identify individuals who might have been exposed to an infectious person. This method was instrumental during disease outbreaks like smallpox \cite{r12} and tuberculosis \cite{r13}. In these cases, meticulous record-keeping and personal follow-ups were key strategies for preventing further spread.

The COVID-19 pandemic spurred unprecedented advancements and the adoption of digital contact tracing tools. Various countries developed mobile applications that utilized GPS and Bluetooth technologies to automate the detection of close contacts \cite{r14}. For example, the TraceTogether \cite{r15} application in Singapore and Apple and Google's joint exposure notification application GAEN \cite{r16} were pivotal in scaling up contact tracing to large populations. These applications could notify users if they had been near someone who tested positive for COVID-19, thereby facilitating a quicker self-isolation or testing response. Though useful in theory, in reality too many false positive alerts resulted in less overall public uptake. 

\subsection{Targeted Mitigation Strategies}
Researchers aimed to measure the efficacy of targeted mitigation strategies such as enacting targeted testing \cite{r17}, quarantines \cite{r18}, and vaccinations \cite{r19}. To do so, they put effort into identifying significant players within the transmission dynamics, such as superspreaders and transmission bottlenecks, through analyzing network metrics \cite{r20}. Most similar to our proposed network metric, Lev et. al. introduced an Infectious Betweenness Centrality that charts the betweenness centrality of a node along the path from an infectious node to a susceptible node \cite{r21}. In contrast, instead of node betweenness, we propose a Infectious Path Centrality metric that describes the nodes that have high betweenness on \textit{paths} connecting one infectious person to another thus identifying the likely path of disease transmission. 

Furthermore, researchers have compared the effectiveness of forward contact tracing, where a person isolates upon testing positive, to bidirectional contact tracing, where an infectious person also retraces their interactions to identify potential unseen infections \cite{r22}. They found that isolating these potential cases can prevent more infections than just isolating those who test positive. Our work aims to provide the first machine learning-based solution that charts the paths of disease transmission.   

\subsection{Graph Learning}

Graph learning has emerged as a crucial discipline within machine learning, primarily due to the ubiquity of graph-structured data across various domains such as social networks, biological networks, or communication systems. The introduction of Graph Neural Networks (GNNs) marked a significant shift towards using deep learning techniques for graph data \cite{r23}. GNNs iteratively update node representations by aggregating features from neighboring nodes, effectively capturing the local graph structure. This paradigm was further extended into Graph Convolutional Networks (GCNs) \cite{r24}, simplifying the graph convolution process and significantly improving the computational efficiency. GCNs have become a cornerstone for numerous applications in link prediction and classification on nodes, edges, and graphs.

Recent work has addressed the challenges of graph learning in the presence of significant class imbalances \cite{r25}. However, the extent of class imbalance in certain applications, such as disease contact tracing, is particularly severe. For instance, in scenarios where the goal is to determine the transmission path of an infection, typically only one edge (or interaction) out of many possible interactions is responsible for the transmission. This results in a class imbalance ratio that is inversely proportional to each node degree when contact tracing.

Given the limitations of handling graph imbalance in such extreme cases, our approach seeks to provide a solution against such profound imbalances. Moreover, disease contact tracing introduces a novel challenge to graph learning: the task of identifying specific paths (or graph traversals) based on transmission dynamics. This necessitates not only managing the severe class imbalance, but also developing techniques capable of accurately tracing the paths of transmission in complex network structures.

In this paper, we build on prior work in disease contact tracing and graph learning to 1) develop a new graph learning framework that learns how to chart the path on infection during a largescale outbreak, and 2) show that bidirectional contact tracing outperforms standard forward contact tracing when trying to mitigate transmission.

\begin{figure*}[ht]
\includegraphics[width=\textwidth]{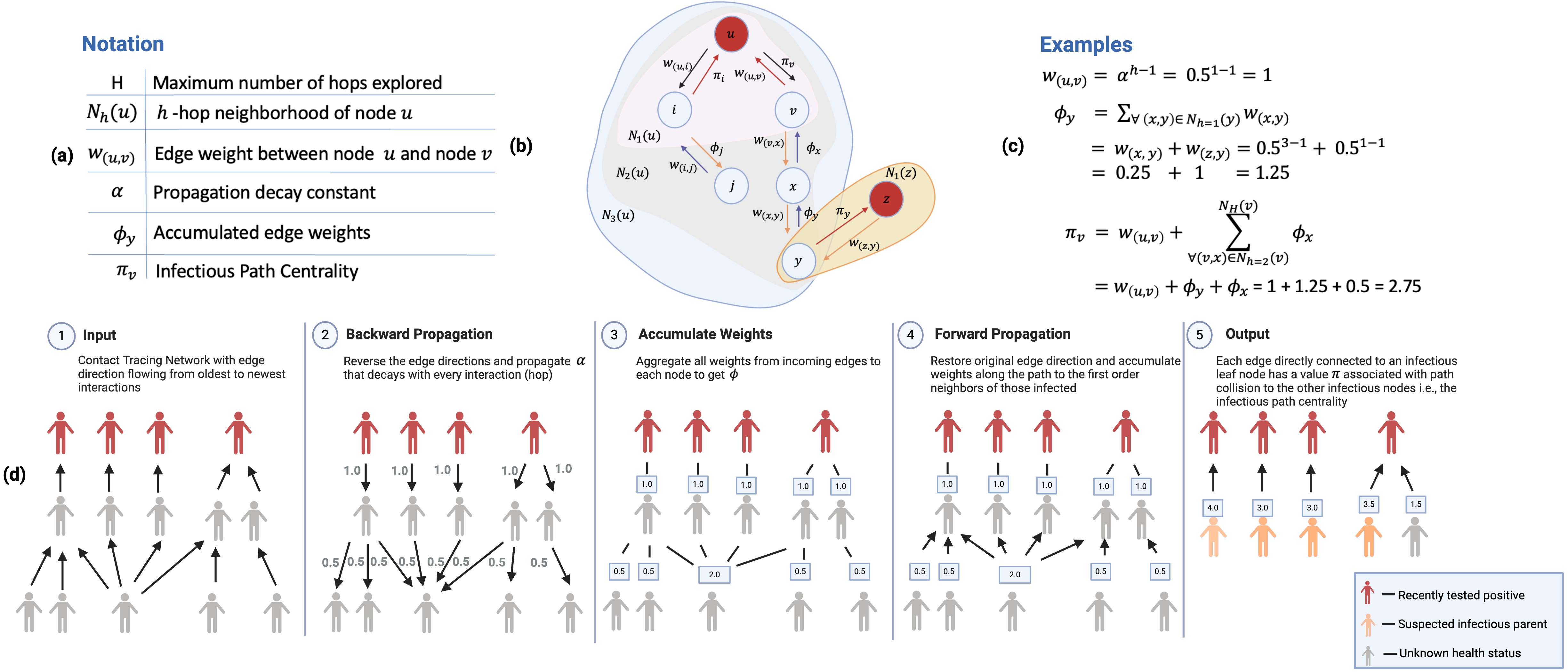}
\caption{   
\textbf{(a)} Notation for equations 2-4. \textbf{(b)} Toy example of a contact tracing network using the notation for our proposed Infectious Path Centrality metric. Nodes $u$ and $z$ just tested positive and are tracing past interactions to identify who is the likely source of infection between nodes $v$ and $i$; we assume node $y$ infected node $z$. $H$ denotes the number of hops (i.e., depth), the Infectious Path Centrality encompasses. $\alpha$ denotes a propagation decay constant that facilitates calculating the weight $w$ of each edge in the $N_h$ ($h$-hop neighborhood). Note that $w$ acts as an attenuating signal originating from node $u$ that sub-samples the larger contact tracing network. The term $\phi_y$ represents the total weights of incoming edges of node $y$ from all paths originating from infectious leaf nodes (i.e., nodes $u$ and $z$). The Infectious Path Centrality $\pi_v$ then quantifies the forward accumulation of all $\phi$ values leading back to $u$'s immediate neighbors (i.e., $N_1(u)$). \textbf{(c)} Example calculations for $w_{(u,v)}$, $\phi_y$, and $\pi_v$. \textbf{(d)} Step-by-step process of calculating Infectious centrality where 1) we input the contact tracing network and 2) reverse the edges to attenuate the weights down to the maximum hop level $H$. 3) Each node accumulates weights $w$ from all incoming edges to get $\phi$. 4) We then restore original edge directions and accumulate $\phi$ back to the first-hop neighbors of the leaf nodes. 5) Next, we add all the $\phi$ from the incoming edges into the first-hop neighbors to create the their respective Infectious Path Centrality value $\pi$. 6) Finally, we normalize all $\pi$ values and use them as features in the edge classification module. By design, if a node exists on paths leading to multiple infectious leafs, the $\pi$ value will be greater. We hypothesize that the (orange) node with the maximum $\pi$ is likely the infectious source.  
 }
\label{fig3}
\end{figure*}

\section{Proposed Approach}
In Fig. \ref{fig2}, we provide a comprehensive overview of our approach. First, we convert individual dwell times from Foursquare mobility data into person-to-person networks. We then apply a network-based compartmental epidemiological model (SEIR) via simulation (step 2). Following this, we map out the path of transmission, creating an accumulative contact tracing network, and compute the Infectious Path Centrality for each neighbor connected to an infectious individual (steps 3 and 4). Using these centrality measures as features, we train a graph edge classifier to identify edges that represent transmission events (step 5). Finally, to evaluate the effectiveness of our model, we conduct new simulations that quarantine nodes based on standard forward tracing and our bidirectional contact tracing method (step 6).

In this section, we describe in detail our approach for network construction, epidemic model, Infectious Path Centrality metric, and mitigation evaluation.

\subsection{Person-to-Person Network}

Given the visits from Foursquare mobility data with venues and dwell times, we construct the initial person-to-person graph $G = (V, E)$ where $V$ is the set of nodes (i.e., persons), and $E$ is the set of edges (i.e., interactions), between them. We define an interaction if two people are at the same venue within the same hour. We construct these graphs for each day in the Foursquare mobility dataset with the time granularity of one hour. We assume an equal chance for any two people to interact within a venue, therefore, within each venue at a given hour the subgraph is fully connected. 

\subsection{Epidemic Model}
Because we lack the ground truth for health labels on each individual within the dataset, we rely on simulating the disease spreading. More precisely, we deploy an agent-based SEIR model \cite{rSEIR} on the population where a node on the contact networks can be one of the four health states, i.e., \{\textit{Susceptible, Exposed, Infectious, Recovered}\}. To introduce heterogeneity, we assign each individual with  immunity $\delta \in [0,1]$ and virality $v \in [0,1]$ values. All nodes start off as \textit{Susceptible} and move to \textit{Incubating} when they have a direct interaction (i.e., edge) with an infectious node that has a virality $v$ greater than their immunity $\delta$ (equation \ref{eq:1}): 

\begin{equation} \label{eq:1}
S_j \rightarrow I_j : v_i > \delta_j 
\end{equation}

After the incubation period (5 days), the individual is considered to be \textit{Infectious} and is assigned a $v$ virality value. After a sickness period (7 days), the individual is considered \textit{Recovered} and cannot be infected again within the testing span of the experiment (30 days).  Note that the immunity threshold, virality value, incubation period, and illness period, are all tunable parameters that could be chosen to simulate a different infectious disease.

\subsection{Contact Tracing Network Construction}
After simulating a disease outbreak on the daily person-to-person networks, we construct a contact tracing graph. This graph aims to emulate the contact tracing performed by a manual contact tracer who calls each infectious individual and queries about their past interactions. To achieve this, for every node that transitions from \textit{Susceptible} to \textit{Incubating}, we add the node's neighborhood to the contact tracing graph where all edges point to the most recently infected node. This forms a directed acyclic graph (tree) where the leaves are all \textit{Infectious} nodes (Fig. \ref{fig2} Step 3).    

\subsection{Infectious Path Centrality}
We abstract the question of \textit{'who infected whom'} to a binary edge classification task to determine whether an interaction contains a transmission event. Because we assume that a person gets infected by one other entity (i.e., person or place), there is a large class imbalance between infectious and non-infectious interactions. 

To account for class imbalance, we take advantage of the transmission dynamics of a disease. As shown in Fig. \ref{fig3}.b, we 
start with the contact tracing network's leaf node $u$ and apply an attenuating signal that has a decay $\alpha \in [0,1]$ with each hop $h$ to act as the edge weights $w_{(u,v)}$ (equation \ref{eq:2})\footnote{All notations relevant to this section are explained in Fig. \ref{fig3}.a}. Note that $\alpha$ gives the importance of each interaction in the $h$-hop neighborhoods and reduces overvaluation of superspreader neighbors. We traverse a maximum of $H$ hops and accumulate the weights of all incoming edges at each node $y$ to get $\phi_y$ (equation \ref{eq:3}). Finally, we reverse the edges to travel from $H$ hops back to the 1-hop neighbors of the infectious node $u$ to accumulate all $\phi$ for all incoming edges resulting in the Infectious Path Centrality value $\pi_v$ (equation \ref{eq:4})

\begin{equation}\label{eq:2}
w_{(u,v)} = \alpha^{h-1}
\end{equation}

\begin{equation}\label{eq:3}
\phi_{y} = \sum_{\forall (x,y) \in N_{h=1}(y)} w_{(x,y)}
\end{equation}

\begin{equation}\label{eq:4}
\pi_{v} = w_{(u,v)} + \sum_{\forall (v,x) \in N_{h=2}(v)}^{N_{H-1}(v)} \phi_x
\end{equation}

Fig. \ref{fig3}.b presents a more extensive toy example to illustrate how calculating Infectious Path Centrality simplifies the contact tracing graph and highlights edges that are part of several transmission paths. To evaluate the efficacy of our proposed metric, we input the Infectious Path Centralities as features into the edge classifier and compare mitigation strategies. 

\subsection{Mitigation}

 We evaluate our mitigation strategies by setting up side-by-side comparison between no-mitigation, forward contact tracing, and bidirectional contact tracing in an 'online' simulation. Starting from interactions in July's person-to-person networks and seed infections, we then test the population of sick individuals with a virality $v > 0.5 $ to simulate testing only symptomatic cases. For each day in the simulation, we then update the contact tracing network and Infectious Path Centralities for each 1-hop neighbor of those who recently test positive. We then use a trained infectious edge classifier using May and June contact tracing network to identify who infected the leaf nodes. In the forward contact tracing, we simply quarantine the individuals who tested positive for the infectious period (7 days). For the bidirectional contact tracing, we go back 5 days (incubation period) and quarantine all of the potential new infections. Finally, we compare the effective reproduction number $R_{t}$ \cite{r26} (that averages the number of new infections caused by one person) of the outbreaks resulting from no mitigation, forward contact tracing mitigation, bidirectional contact tracing mitigation. 

\section{Results}\label{sec2}

In this section, we present our experimental set up, investigate the $H$ hops for the Infectious Path Centrality, perform an ablation study on  $\alpha$, and evaluate a few mitigation strategies.

\subsection{Experimental Set Up}

We built our framework using four months (i.e., May, June, July, August) in 2020, of the Foursquare mobility dataset during various stages of COVID-19 lockdown and reopening in Austin, Texas. For each month, we seed 1\% of the person-to-person networks with infections and form the resulting disease contact tracing network. The statistics for devices captured, number of infections, and contact tracing network nodes and edges are shown in Table 1. 

\vspace{-5pt}
\begin{table}[h] \label{t1}
\caption{Foursquare Sample Size Statistics for 2020}
\centering
\begin{tabular}{ |c|c|c|c|c|  }
 \hline 
 Month & Devices & Infections & Nodes & Edges \\
 \hline
  May & 37,049 & 17,075 & 20,095 & 116,615  \\ 
  June & 37,039 & 20,036  & 21,974 & 154,548\\
  July & 36,347 & 17,644 & 19,899 & 124,821 \\ 
  August & 47,598 & 28,082 & 31,073 & 244,357 \\
 \hline
\end{tabular}
\vspace{-5pt}
\end{table}

\begin{figure}[!b]
\includegraphics[width= \linewidth]{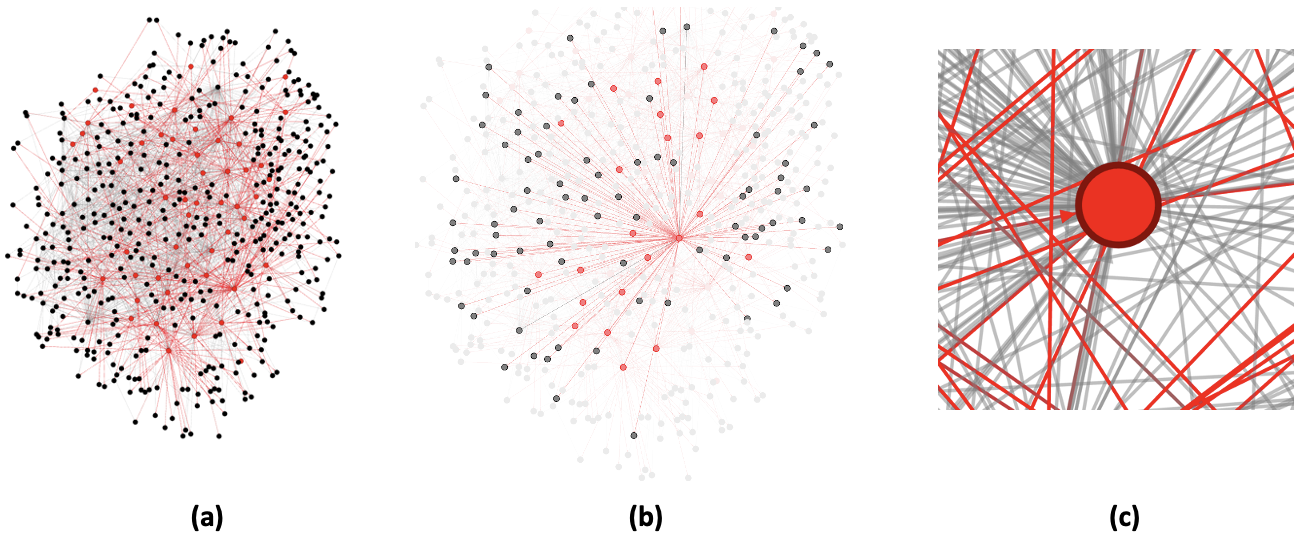}
\caption{   
(a) 500 node sub-sample of a contact tracing network. Red edges signify a transmission event where the parent node infects the child node. (b) portrays an ego-network view of a leaf node that has recently tested positive. Their ego-network consists of all contacts made on the day of infection (incoming edges), as well as those they have infected (outgoing red edges). (c) depicts a zoomed in view of an infectious node where there is only one incoming red edge (i.e., source of infection), and many outgoing red edges (parent of infections). There are also many incoming grey edges that signify interactions on the day of infection that were not transmission events. 
 }
\label{fig4}
\end{figure}

\begin{figure}[h]
\centering
\includegraphics[width=.75 \linewidth]{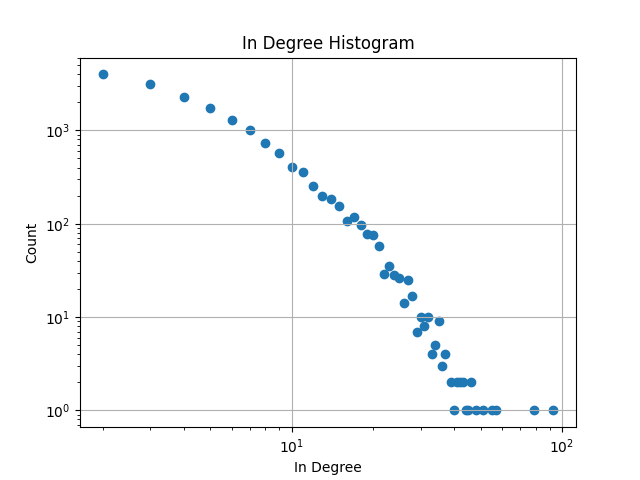}
\caption{   
The in-degree histogram in log-log scale shows that the contact tracing network is scale-free. This means that few nodes have many incoming edges, while most nodes have few. The class imbalance for identifying the incoming transmission edge is proportional to the in-degree which makes training an edge classifier largely unbalanced.}
\label{fig5}
\end{figure}

We visualize the contact tracing network in Fig. \ref{fig4}(a-c) to show the relationship between node infections and edge infections. The nodes in red have contracted the virus, and the edges in red chart the path of infection. Every node has only one red incoming edge Fig. \ref{fig4}.c which signifies that every node is only infected by one person. Fig. \ref{fig5} shows the node degree distribution for all the months captured in the experiment. Given the heavy tailed distribution, we can conclude that the directed contact tracing network is scale-free graph where there exists few superspreaders.

\subsection{Infectious Path Centrality}
\begin{figure}[h]
\centering
\includegraphics[width= .71 \linewidth]{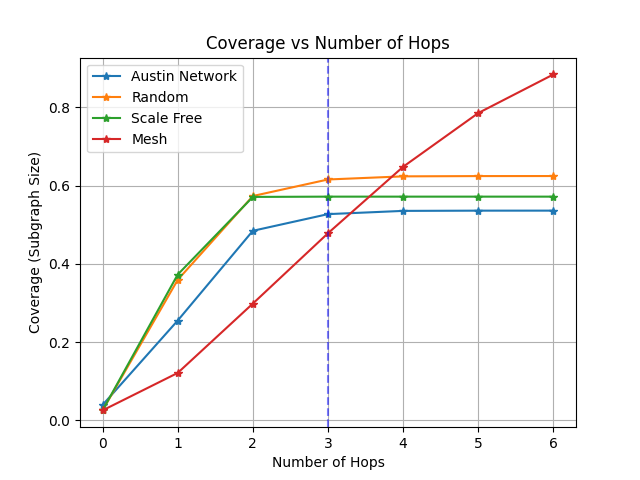}
\caption{   
Comparison of percentage of nodes explored between topologies according to hop-depth. Each network consists of 20,000 nodes and ~116,000 edges to be comparable to the Austin contact tracing network. Of note, the contact tracing network is a DAG structure whereas, the scale-free and random graphs are undirected. The scale free network coverage plateaus at 2 hops and both the Austin contact tracing and random networks plateau at 3 hops. In contrast, the mesh network coverage increases with the number of hops. }
\label{fig6}
\end{figure}

After constructing the contact tracing, we calculate the Infectious Path Centralities for each positive node's 1-hop neighbor, and analyze the maximum number of hops $H$ (i.e., depth) needed in order train the edge classifier well. We first investigate the amount of graph covered by traversing each hop on different graph toplogies. In Fig. \ref{fig6}, we compare the contact tracing graph pulled from a disease simulation on Austin in May of 2020, to scale free, random, and mesh networks of equal size ($20,000$ nodes). 

We start from a sample of 500 nodes on each graph, traverse to each \textit{h}-hop neighborhood and keep track of how many unique nodes are visited. We find that for all cases except the mesh graph, traversing to the 3-hop neighborhood samples the largest subset of the graph (Fig. \ref{fig6}). Though the random, scale free, and Austin contact tracing graphs all have an average path length of around 6, this coverage suggests that the sample nodes are roughly 3 hops away from a superspreader. In contrast, the mesh covers more nodes with every hop.  

\begin{figure}[h]
\centering
\includegraphics[width= .75 \linewidth]{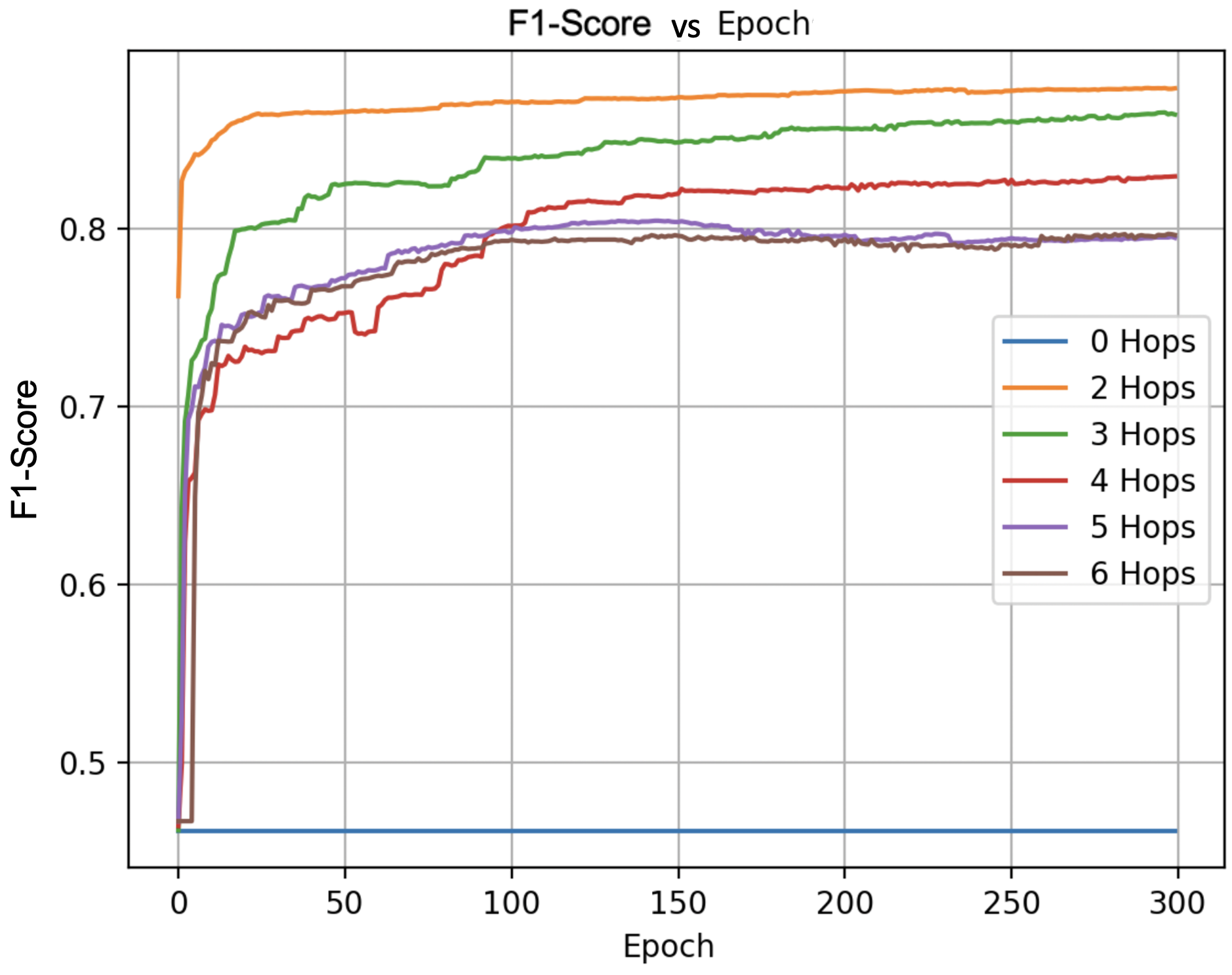}
\caption{   
Capturing F1-score from training edge classifier over 300 epochs while using different number of hops. 0 hops signifies performing edge classification without the Infectious Path Centrality metric. The contact tracing graph is taken from Austin mobility interactions from May, 2020. The Infectious Path Centrality metric calculated by traveling 2-hops away from the leaf node gets the highest F1-score of 0.87 after 300 epochs. 
 }
\label{fig7}
\end{figure}

To further investigate, we train the edge classifier on different maximum hop depths $H$ to compare the F1-scores in Fig. \ref{fig7}. We use the F1 metric \cite{r27} to take into account the class imbalance rather than accuracy since most of the edges are not transmission events. We can see the edge classification model achieves the highest F1-score when traversing to 2-hop neighbors (orange curve). Note that this is one-hop less than achieving the largest coverage (3-hops). Perhaps this is because at 3-hops the Infectious Path Centrality metric has a hard time saturating as there are too many collisions to differentiate the paths.

\begin{figure}[h]
\centering
\includegraphics[width= .75 \linewidth]{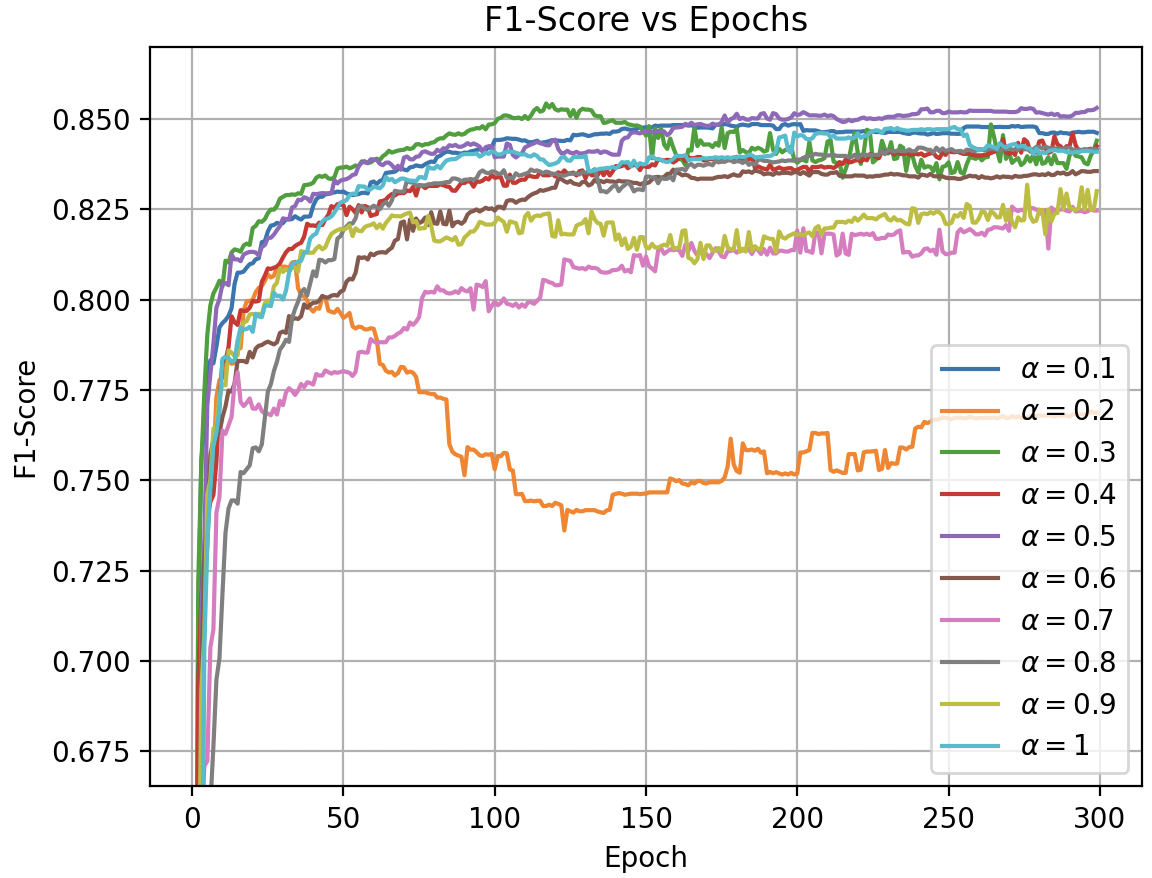}
\caption{\textbf{[Ablation study]} Comparison of $\alpha$ value used on Austin contact tracing network for May of 2020 using $h=2$. For example, an $\alpha=0.1$, means that the propagation signal diminishes quickly between $h$ hops. Decaying the signal by $\alpha=0.5$ yields the highest F1-score of 0.85. }
\label{figalpha}
\end{figure}

\textbf{[Ablation study]} In addition, we investigate various values of $\alpha$ to determine how important we should weigh each $h$-hop neighborhood; we find that $\alpha=0.5$ yields the highest F1-score (Fig. \ref{figalpha}).

\begin{figure}[h]
\centering
\includegraphics[width= .75 \linewidth]{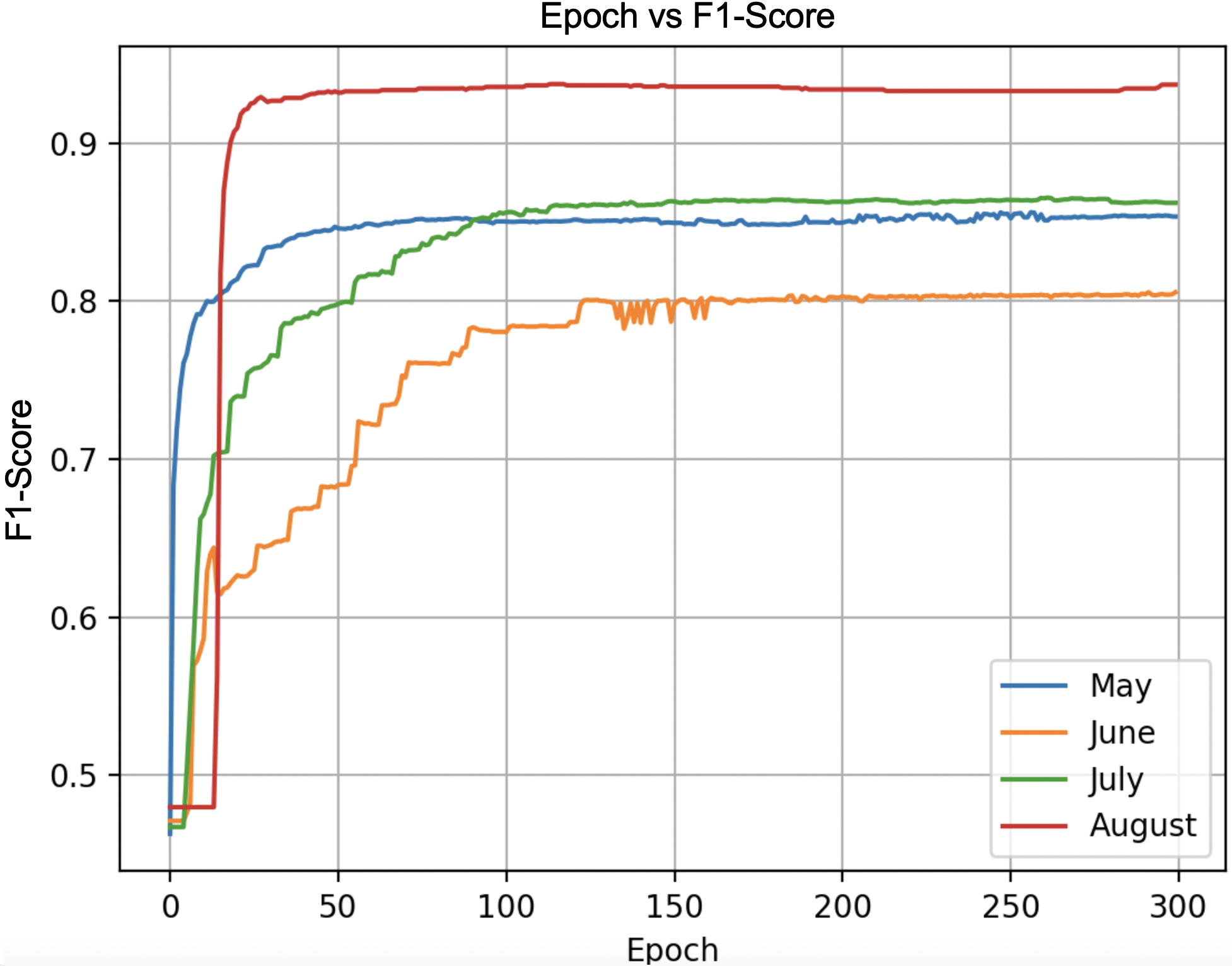}
\caption{   
Comparing training edge classifier on contact tracing networks from Austin in May, June, July, and August in 2020. August has the most nodes (31,073) and edges (244,357) resulting in the largest sub-sample for training. The edge classifier achieves an F1-score of 0.94 after training for 300 epochs on August. In contrast, July has the smallest network with a 0.81 F1-score.    
 }
\label{fig8}
\end{figure}

As such, we train all of the months using $h$ = 2 with $\alpha = 0.5$ in Fig. \ref{fig8} where F1-scores range between 0.81 and 0.94. After determining the number of hops that the flow metric should traverse, and training the edge classification model, we investigate the efficacy of bidirectional contact tracing. 

\begin{figure}[h]
\centering
\includegraphics[width=.75 \linewidth]{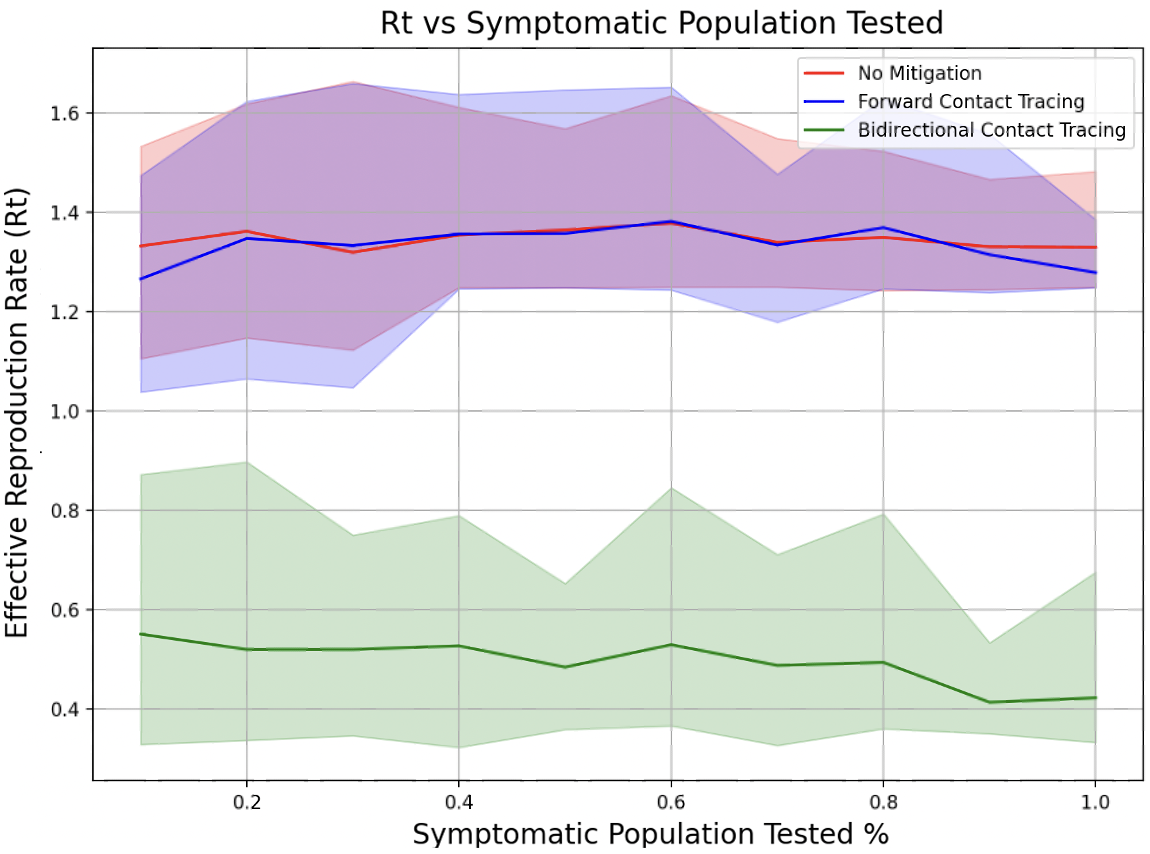}
\caption{ Comparison of effective reproduction number $R_t$ across 20 simulation runs for three scenarios: no mitigation measures, forward contact tracing, and bidirectional contact tracing, varying the percentage of symptomatic cases tested. The mitigation strategies are implemented starting on the eighth day, and $R_t$ represents the average number of secondary infections per case by the end of each simulation. The shades with lighter colors represent the minimum and maximum $R_t$ among the 20 simulations.
}
\label{fig9}
\end{figure}

\subsection{Mitigation}

 We utilize an edge classifier trained on contact tracing networks from May and June and apply 'online' mitigation to July and August. As described in the approach, we test newly symptomatic individuals and calculate the Infectious Path Centralities by traversing to their $H=2$-hop neighbors. We then update the contact tracing network and classify the edges of the 1-hop neighbors to gather the parent infections. From there, we test whether quarantining those who tested positive (forward contact tracing), or retroactively quarantining all neighbors of those who infected the positive individuals (bidirectional contact tracing) lowers the disease reproduction rate compared to the unmitigated population. 
 
 In each case, we seed the infections with the same 1\% of individuals so that the infection comes from the same starting point. We implement the mitigation techniques starting on the eighth day to ensure that the disease has propagated. Fig. \ref{fig9} shows a comparison of the effective reproduction number at the end of each simulation for different percent of population tested. For example, when only 1\% of symptomatic cases (i.e., virality $v > 0.5$) is tested, the unmitigated population has an average effective reproduction rate of 1.33 while the forward tracing has a rate of 1.27 and bidirectional contact tracing has 0.54. Most notably, regardless of the percent of population tested, we see a dramatic decrease in effective reproduction rate between bidirectional contact tracing and unmitigated populations (71\%) and forward contact tracing (54\%). This is in contrast to only using forward contact tracing which decreases the effective reproduction rate by only 16\%. Intuitively, this is because when performing forward contact tracing while only testing a percentage of the symptomatic cases, many transmission paths will go untested which results in the continuation of viral propagation. When 100\% of the symptomatic individuals are tested, the bidirectional contact tracing mitigation strategy results in a effective reproductive rate of 0.42 which means that, on average, every ten infected people produce roughly four infectious offspring, thereby significantly slowing down the outbreak.

\section{Conclusion}

In this paper, we have presented a framework to automate bidirectional contact tracing using Foursquare mobility data. We have formulated the transmission path identification problem on contact tracing networks as graph learning edge classification that determines whether an edge is a transmission event. We have also proposed a new network metric, Infectious Path Centrality, that describes the centrality of a node along the path from two infectious nodes. 

Our proposed metric solves the class imbalance problem on transmission path identification, as it sub-samples the nodes along the $H$-hop neighborhoods. Moreover, unlike manual contact tracing, our metric performs better in scenarios of widespread community transmission. Through training, our edge classification model achieves an F1-score of 0.94; when used to perform bidirectional contact tracing, the $R_t$ decreases by $71\%$ compared to unmitigated populations. 

Our work is limited by the fact that we do not have the ground truth for the true transmission dynamics, as there is no public dataset that contains large scale interactions, as well as health labels. Future work includes extending our analysis to other cities and communities of various size (i.e., New York City vs St. Louis) to evaluate the viability of our proposed metric. Furthermore, we intend to investigate the robustness to imperfect quarantines with varying degree of compliance, and inaccurate testing (i.e., false positives and/or false negatives). 

Taken together, our work is an important step towards automating disease contact tracing to help mitigate the next unknown viral outbreak. 


\printbibliography

\end{document}